\documentstyle[psfig]{mn}

\newcommand{\kms}{\mbox{${\;{\rm km\,s^{-1}}}$}}
\newcommand{\A}{\mbox{$\,\mathrm{\AA}$}}
\newcommand{\ee}[1]{\mbox{${} \times10^{#1}$}}
\newcommand{\gs}{\mbox{$\,{\mathrm g}\,{\mathrm s}^{-1}$}}

\newcommand{\Msolar}{\mbox{${\; {\rm M_{\sun}}}$}}
\newcommand{\HeI}{He$\,${\sc i}}
\newcommand{\ew}{\mbox{$W_\lambda$}}

\newcommand{\cir}{Cir~X-1}
\newcommand{\xte}{{\em RXTE}}

\newcommand{\rb}[1]{\raisebox{1.5ex}[0pt]{#1}}

\psfigurepath{/home-theory/helenj/SCR/Figures/CirX-1/}

\title[Variability of optical spectra of Cir X-1] {Secular and orbital
  variability of Cir X-1 observed in optical spectra}

\author[H. M. Johnston et al.]{Helen M. Johnston,$^1$\thanks{E-mail:
  H.Johnston@physics.usyd.edu.au; kw@mssl.ucl.ac.uk; rpf@astro.uva.nl;
  J.Cullen@physics.usyd.edu.au}, Kinwah Wu$^{1,2}$, Rob Fender$^3$ and
  Jason G. Cullen$^1$ \\  
  $^1$Research Centre for Theoretical Astrophysics, School of Physics,
  University of Sydney, NSW 2006, Australia; \\
  $^2$Mullard Space Science Laboratory, University College London,
  Holmbury St Mary, Dorking, Surrey RH5 6NT; \\
  $^3$Astronomical Institute `Anton Pannekoek', University of
   Amsterdam, Kruislaan 403, 1098 SJ Amsterdam, The Netherlands }

\date{Received: } 

\begin{document}

\maketitle

\begin{abstract}
     We have observed variations in the optical emission lines from
     the X-ray binary Circinus~X-1. These variations may be attributed
     both to orbital variations and to long term secular changes in
     line strength. We have detected double-peaked H$\alpha$\ emission
     lines on two occasions, providing the first direct evidence for
     an accretion disk in the system. The separation of the peaks was
     different on the two occasions, suggesting that the disk might
     have a different size.  The equivalent width of the emission
     lines dropped by more than a factor of three between 1999 and
     2000; this continues the trend seen in earlier data, so that the
     H$\alpha$\ equivalent width has now declined by a factor of
     twenty since 1976.  The emission lines do not appear to show
     signature of orbital motion, except for the data taken near phase
     0, which show a significant velocity shift.
     
     We have observed an absorption component to the \HeI\ lines on
     one occasion. We suggest that, unlike the P~Cygni profiles seen
     in X-ray spectra, this absorption does not arise in the
     accelerating zone of a radiatively driven wind. Instead, the
     absorption arises in material previously ejected from the system.
     It was only seen on this one occasion because the strength of the
     emission line had dropped dramatically.
\end{abstract}

\begin{keywords}
binaries: spectroscopic -- stars: individual: Cir X-1 -- stars: X-rays
\end{keywords}

\section{Introduction}
\label{sec:intro}

\cir\ is a highly unusual X-ray binary, whose nature has been a puzzle
for many years. Since the early 1970s, Cir~X-1 has shown erratic X-ray
properties: its light curve differed dramatically each time it was
observed.  Nevertheless, there is a periodic modulation of 16.6~d
\cite{khbs76}, which is believed to be the orbital period of the
binary.  The compact star in Cir X-1 is a neutron star, inferred from
the Type I X-ray bursts observed in a brief episode (Tennant, Fabian
\& Shafer 1986\nocite{tfs86b}). Further support is provided by the
Z-source behaviour seen recently in X-ray timing data from the {\em
  Rossi XTE} satellite (Shirey, Bradt \& Levine 1999\nocite{sbl99}).

The radio counterpart of Cir X-1 is located $25'$\ from the centre of
the supernova remnant G321.9$-$0.3, and is apparently connected to the
remnant by a radio nebula \cite{hkl+86}.  It shows flares at the same
period as the X-ray modulation \cite{hjm+78}.  Cir X-1 has two
arcminute-scale radio jets \cite{schn93}, and an arcsecond-scale
asymmetric jet suggests the presence of relativistic outflow from the
source \cite{fst+98}.  The close association of Cir~X-1 with the
supernova remnant suggests that the system may be a young ($< 10^5$~y
old) runaway system from a supernova explosion \cite{schn93}.

\begin{table*}
\begin{minipage}{\textwidth}
\caption{Journal of observations of Cir~X-1. The columns show the UT date
  of observation, the Julian date of the midpoint of the observation,
  the telescope and instrument used, the mean phase of the
  observation, the total exposure time, and the wavelength range and
  resolution of the spectra, as measured from the arc lines. The phase
  was calculated according the ephemeris of Stewart et al. (1991).
  \label{tab:obs-log}}\nocite{snp+91}
\end{minipage}
\begin{tabular}{lcl c r r@{--}l c}
\hline
UT Date & JD & Instrument & Phase & \multicolumn{1}{c}{$t_{\mathrm exp}$} &
    \multicolumn{2}{c}{Wavelength} & Resolution \\
        &    &            &       & \multicolumn{1}{c}{(s)}               &
    \multicolumn{2}{c}{range (\AA)} & (\AA) \\
\hline
1999 Jul 11 & 2451371.045 & 2.3m + DBS & 0.880 & 18900 & 5600 & 7500 & 2.7 \\
1999 Aug 20 & 2451410.910 & 2.3m + DBS & 0.292 & 5400  & 5400 & 7330 & 2.1 \\
1999 Aug 21 & 2451411.870 & 2.3m + DBS & 0.350 & 1800  & 5400 & 7330 & 2.1 \\
1999 Aug 22 & 2451412.901 & 2.3m + DBS & 0.411 & 7200  & 5400 & 7330 & 2.1 \\
2000 May 16 & 2451681.045 & AAT + RGO  & 0.622 & 28800 & 5660 & 7000 & 1.3 \\
2000 May 22 & 2451687.063 & 2.3m + DBS & 0.985 & 28800 & 5570 & 7000 & 2.4 \\
2000 Jul 12 & 2451738.096 & 2.3m + DBS & 0.073 & 12600 & 5580 & 7510 & 4.8 \\
2000 Jul 13 & 2451739.013 & 2.3m + DBS & 0.129 & 19800 & 5508 & 7510 & 4.5 \\
\hline
\end{tabular}
\end{table*}

The optical counterpart to Cir~X-1 was identified as a highly-reddened
star with strong H$\alpha$\ emission \cite{wmw+77}. This object was
later shown to consist of three stars within a radius of 1\farcs5, the
southernmost of which is the true counterpart \cite{mon92,dsh93}. In
1997, we obtained spectra of Cir~X-1 near apastron\footnote{Here and
  throughout, we use the words ``periastron'' and ``apastron'' to
  refer to phase 0 and phase 0.5 respectively, as calculated from the
  timing of the radio flares \cite{snp+91}.}, using the
Anglo-Australian Telescope (Johnston, Fender \& Wu 1999, hereafter
Paper I)\nocite{jfw99}. We detected an asymmetric H$\alpha$ line which
can be decomposed into two components, a narrow one and a broad
component which is blue-shifted with respect to the narrow component.
Comparison with archival spectra from the AAT and from HST showed that
an asymmetric emission line has been present for the past twenty
years. No previous observations have seen any evidence of an accretion
disk.

In this paper, we report on new optical spectroscopic observations of
\cir\ obtained during 1999 and 2000.

\section{Observations and data reduction}
\label{sec:obs-data-reduct}

\cir\ was observed on multiple occasions during 1999 and 2000, using
the Double Beam Spectrograph (DBS) on the ANU 2.3-m telescope, and the
RGO Spectrograph on the 3.9-m AAT. A complete log of observations is
given in Table~\ref{tab:obs-log}.

The 2000 May 16 observations were carried out with the RGO
spectrograph in combination with the MITLL3 chip and a grating with
1200~grooves$\;{\mathrm mm}^{-1}$, resulting in a dispersion of
$1.3\A\;\mathrm{pixel}^{-1}$. All other observations were
carried out using the DBS with the standard SITe $1752 \times 532$\ 
CCD with a 600~grooves$\;{\mathrm mm}^{-1}$\ grating.  The very red
colour of the source means that no significant signal was detected in
the blue arm of the DBS. Series of 1800$\;$s object exposures were
interspersed with CuAr arc-lamp exposures.  A slit width of 1.5 or
2~arcsec was used, and the slit was always oriented north-south so
that both \cir\ and star 2 of Moneti (1992)\nocite{mon92} were in the
slit. The spatial scale on the detector was 0\farcs91 for the DBS and
0\farcs48 for the RGO, which means the spectrum of \cir\ was confused
with that of star~2; keeping the same slit orientation means that at
least this contribution was constant. Because our previous work
(Paper~I)\nocite{jfw99} had shown no significant features associated
with star~2, we have not attempted to remove its contribution from the
spectrum.

The phases of the observations are shown in Table~\ref{tab:obs-log},
as calculated from the ephemeris of Stewart et al. \shortcite{snp+91},
based on the timing of the radio flares.  It can be seen that, while
there is moderately good coverage of a range of orbital phases, there
is a year separating the earliest spectra from the last spectra, so
there are potential secular variations which may confuse any features
arising from the orbital variation.

The {\sc iraf} software suite was used to remove the bias and
pixel-to-pixel gain variations from each frame.  As we had multiple
consecutive observations of the same object, cosmic ray events were
removed using the technique described by Croke \shortcite{cro95}, as
implemented in {\sc figaro}. The wavelength calibration was performed
in {\sc iraf} by fitting a low-order polynomial to the arc line
wavelengths as a function of pixel number; the rms scatter of the fits
was typically better than 0.1~pixel.

\section{Results}
\label{sec:Results}

The spectra all show a strong H$\alpha$ emission line; the shape
however changed significantly between the various observations. We can
see both long-term secular variation, in the equivalent widths of the
lines, and variations which seem to be correlated with the orbital
phase of the observations, in the line-profile morphology.
Unfortunately, with sparse orbital coverage, it is difficult to
disentangle the latter from the former.  Only with observations within
a single orbit, and repeated observations at the various orbital
phases, will we be able to be certain about some of the phenomena we
describe here.  However, with that caveat in mind, we will describe
the data at hand.

\begin{table*}
\caption{Fits to the emission line profiles. Column 2 shows the mean
  phase of the observation, column 3 shows the total equivalent width
  of the H$\alpha$\ line $W_{\lambda,t}$, measured by direct summation
  of pixels in the emission line, and columns 4 and 5 show the
  equivalent width of the \HeI\ $\lambda 6678$\ and $\lambda 7065$\ 
  lines, measured from Gaussian fits to the line profiles (see
  Sect.~\protect{\ref{sec:He-I-lines}}). The \HeI\ lines in the 2000
  May 16 spectra show P~Cygni profiles; the equivalent widths listed
  are those of the emission component. Columns 6--11 show the results
  of Gaussian fits to the H$\alpha$\ line. The spectra for 1999 July
  11 and 2000 May 16 were fit with three components, the other spectra
  with two.  The fits are shown in Figure~1.}\label{tab:fits}\addtolength{\tabcolsep}{-1pt}
\begin{tabular}{@{}lcc r@{$\,$}c@{$\,$}l r@{$\,$}c@{$\,$}l r@{$\,\pm\,$}l
  r@{$\,\pm\,$}l r@{$\,\pm\,$}l r@{$\,$}c@{$\,$}l r@{$\,$}c@{$\,$}l
  r@{$\,$}c@{$\,$}l} 
\hline
 & & $W_{\lambda,t}$ & \multicolumn{3}{c}{$W_\lambda$} & 
\multicolumn{3}{c}{$W_\lambda$} & 
\multicolumn{6}{c}{H$\alpha$\ narrow component(s)} &
        \multicolumn{9}{c}{H$\alpha$\ broad component} \\
UT Date & Phase & 
H$\alpha$ & 
\multicolumn{3}{c}{$\lambda 6678$} & 
\multicolumn{3}{c}{$\lambda 7065$} & 
\multicolumn{2}{c}{velocity} &
\multicolumn{2}{c}{FWHM} & \multicolumn{2}{c}{$W_{\lambda}$} &
\multicolumn{3}{c}{velocity} & \multicolumn{3}{c}{FWHM} &
\multicolumn{3}{c}{$W_{\lambda}$} \\
        &       &  (\AA) & 
\multicolumn{3}{c}{(\AA)} & \multicolumn{3}{c}{(\AA)} & 
\multicolumn{2}{c}{(${\rm km\,s^{-1}}$)} &
\multicolumn{2}{c}{(${\rm km\,s^{-1}}$)} &
\multicolumn{2}{c}{(\AA)} & 
\multicolumn{3}{c}{(${\rm km\,s^{-1}}$)} &
\multicolumn{3}{c}{(${\rm km\,s^{-1}}$)} &  
\multicolumn{3}{c}{(\AA)} \\
\hline
& & & & & & & & & $-$104 & 10 & 270 & 16 & 21 & 2 \\
\rb{1999 Jul 11} & \rb{0.880} & \rb{81} & 
\rb{3.5} & \rb{$\pm$} & \rb{0.4} & \rb{4.7} & \rb{$\pm$} & \rb{0.3} &
270 & 10 & 400 & 15 & 40 & 1.5 &
\rb{$-600$} & \rb{$\pm$} & \rb{40} & 
\rb{550} & \rb{$\pm$} & \rb{70} & 
\rb{9.1} & \rb{$\pm$} & \rb{1.1} \\
1999 Aug 20 & 0.292 & 80 & 4.6 & $\pm$ & 0.9 & 6.7 & $\pm$ & 1.0 &
  261 &  8 &  490 & 20 & 47 & 4 & 
  $-$90 & $\pm$ & 60 & 1010 & $\pm$ & 60 & 27 & $\pm$ & 4 \\
1999 Aug 21 & 0.350 & 79 & 1.7 & $\pm$ & 1.0 & 4.4 & $\pm$ & 0.8 &
  350 & 16 &  290 & 40 & 24 & 6 &
  180 & $\pm$ & 30 &  675 & $\pm$ & 50 & 51 & $\pm$ & 6 \\
1999 Aug 22 & 0.411 & 83 & 3.2 & $\pm$ & 0.8 & 3.9 & $\pm$ & 0.6 &
  330 & 10 &  320 & 24 & 30 & 5 &
  1040 & $\pm$ & 30 &  690 & $\pm$ & 30 & 49 & $\pm$ & 5 \\
& & & & & & & & & $-$30 & 5 & 210 & 40 & 2.7 & 0.7 \\
\rb{2000 May 16} & \rb{0.622} & \rb{12.2} & 
  \rb{0.28} & \rb{$\pm$} & \rb{0.08} & \rb{0.66} & \rb{$\pm$} & \rb{0.09} &
  200 & 5 & 170 & 30 & 3.1 & 0.7 &
  \rb{$-$420} & \rb{$\pm$} & \rb{30} & \rb{630} & \rb{$\pm$} & \rb{45} &
  \rb{6.6} & \rb{$\pm$} & \rb{0.9} \\
2000 May 22   & 0.985 & 10.1 & 1.4 & $\pm$ & 0.2 & 0.3 & $\pm$ & 0.1 &
  87 & 15 &  270 & 60 & 1.5 & 0.5 &
  44 & $\pm$ & 18 &  1000 & $\pm$ & 50 & 9.5 & $\pm$ & 0.5 \\
2000 Jul 12   & 0.073 & 25.0 & 2.0 & $\pm$ & 0.5 & 1.9 & $\pm$ & 0.3 &
  307 & 80 &  580 & 60 & 11 & 15 &
  $-$20 & $\pm$ & 400 &  740 & $\pm$ & 230 & 10 & $\pm$ & 15 \\
2000 Jul 13   & 0.129 & 26.3 & 2.6 & $\pm$ & 0.5 & 2.3 & $\pm$ & 0.4 &
  290 & 40 &  440 & 40 & 13.8 & 3 &
  $-$115 & $\pm$ & 70 & 460 & $\pm$ & 80 &  8.0 & $\pm$ & 3 \\
\hline
\end{tabular}
\end{table*}

\subsection{H$\alpha$\ line shape}
\label{sec:Halpha-line-shape}

As was seen in previous observations (Paper~I), the H$\alpha$\ line
shows multiple components.  In particular, the spectrum taken on 1999
July 11 shows a clear double-peaked profile. Such profiles are well
known in the optical spectra of soft X-ray transients and cataclysmic
variables, where they are usually taken to indicate the presence of an
accretion disk \cite{sma81}. A natural interpretation for the
double-peaked lines in our spectra would therefore be that there is an
accretion disk present, although other possible interpretations are
discussed below (\S~\ref{sec:Detection-accretion-disk}).  This is the
first time evidence of the accretion disk has been seen in the optical
spectra of Cir X-1.

\begin{figure}
     \centerline{\psfig{figure=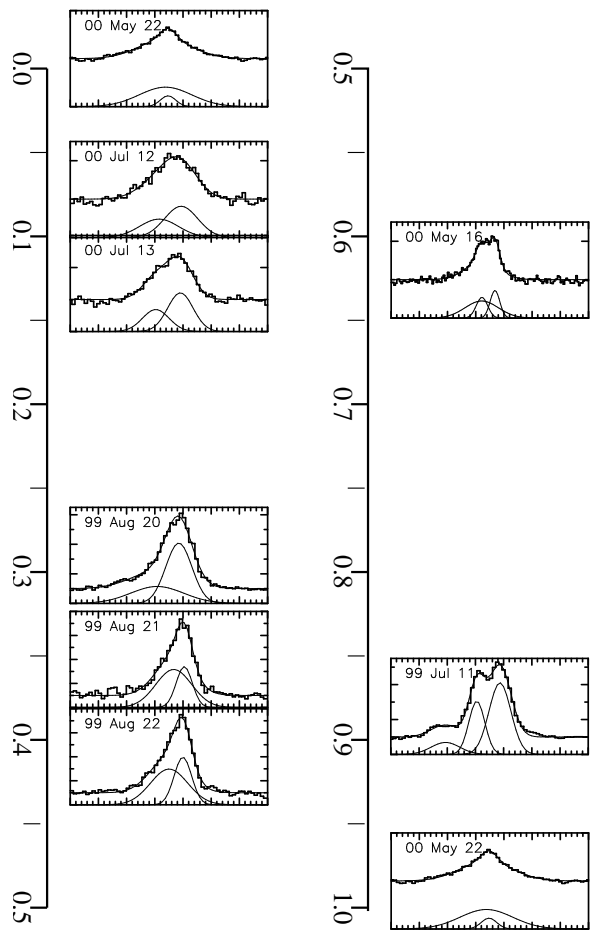,width=7.7cm,clip=t}}
\caption{Line profiles of H$\alpha$, showing the Gaussian fits to the
  lines and their sum. The spectra have been normalised by a
  polynomial fit to the continuum. The spectra are shown in order of
  phase, with the spectrum taken at phase 0 (1999 May 22) repeated at
  phase 1 for clarity. The spectrum at phase 0 is symmetric, with a
  broad component on the blue wing appearing at phases 0.1--0.5. At
  phase 0.6 the line appears to be double-peaked (or flat-topped?),
  while clear double peaks are seen at phase 0.9.
  }\label{fig:Halpha-fits}
\end{figure}

We fit Gaussian profiles to the H$\alpha$ line, using the {\tt
  specfit} package in {\sc iraf} \cite{kri94}. After normalising each
spectrum by a low-order polynomial fit to the continuum, we fit either
two or three gaussians to the H$\alpha$\ line.  The spectrum taken on
1999 July 11 clearly showed a double-peaked profile, in addition to
the broad blue-shifted component, so we fit three gaussians to this
spectrum.  The reduced $\chi_{\nu}^2$\ for $\nu=85$\ degrees of
freedom changes from $\chi_{\nu}^2 = 4.9$\ for a two-gaussian fit to
$\chi_{\nu}^2 = 1.1$\ for three gaussians.  The spectrum from 2000 May
16 also required an extra narrow component to achieve a good fit
($\chi_{\nu}^2 = 1.52$\ for a two-gaussian fit, $\chi_{\nu}^2 = 1.23$
for three gaussians).  The other spectra were fit with two gaussians.
The details of the fits are listed in Table~\ref{tab:fits}.

The data and the fits are shown in Figure~\ref{fig:Halpha-fits}, with
the spectra arranged in order of phase. Despite the spectra having
been taken over a span of more than a year, there appears to be a
clear trend of line shape with orbital phase. The spectrum taken at
phase 0 (2000 May 22) is the only completely symmetric profile.  At
small phases, the line profile begins to show asymmetry in the line.
Sometime after phase 0.4, the line shows a second peak in addition to
the broad component (2000 May 16), which persists for the second half
of the orbit (1999 July 11). We have fit two narrow components to the
line at phase 0, although it is possible that the profile is
flat-topped instead of intrinsically double-peaked. We will discuss
the implications of this in
Section~\ref{sec:Detection-accretion-disk}.

In Paper~I we presented five archival spectra from the AAT and one
from HST; we can re-examine these spectra for evidence of the same
pattern in the line profiles.  Three spectra (1976 May, 1978 August
and 1997 June) were taken near apastron, at phases 0.54, 0.52 and
0.51; none of them shows a double-peaked profile. The spectra taken in
the first half of the orbit all show an asymmetry on the blue wing; no
spectra were taken during the second half of the orbit (phases
0.6--0.9), so we do not know whether the double-peaked line has been
consistently present. The spectrum taken near periastron (1977
September, phase 0.95) does not show a symmetric profile; the line is
still asymmetric on the blue wing. Thus we can say very little about
the persistence of the pattern we find in our spectra from 1999--2000.

\subsection{Secular variation in H$\alpha$ equivalent width}
\label{sec:Secular-variation}

\begin{figure}
     \centerline{\psfig{figure=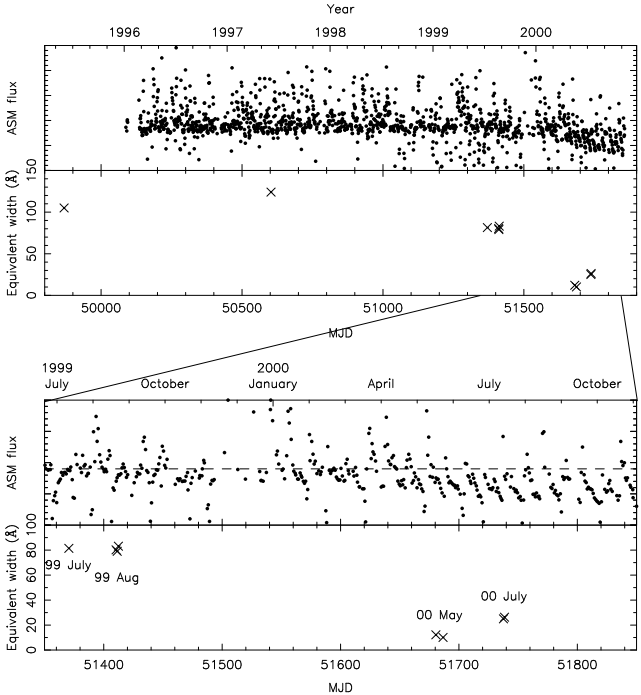,width=8.5cm,clip=t}}
     \caption{Variation of the X-ray flux and equivalent width of the
       H$\alpha$ line with time.  The top section of each panel shows
       the 1-day average ASM flux measured by XTE, the lower section
       shows the equivalent width of the H$\alpha$ line. The entire
       RXTE/ASM light curve for Cir X-1 is shown in the top panel,
       while an expanded version of the light curve since 1999 is
       shown in the bottom panel. The ``baseline'' X-ray flux, between
       flares and dips, began to decline from a value of about
       1.2~Crab (shown by the dashed line) to 0.7~Crab by 2000 July
       (MJD 51740). The H$\alpha$\ equivalent widths for 1995 (phase
       0.18) and 1997 (phase 0.51) are from Paper~I. The drop is
       independent of phase. }\label{fig:eqw}
\end{figure}

The equivalent width \ew\ of the H$\alpha$ line is shown in
Table~\ref{tab:fits} and its evolution is plotted in
Figure~\ref{fig:eqw}. The equivalent width dropped dramatically
between 1999 (MJD 51370--51412) and 2000 (MJD 51680--51738), from
80\A\ to 25\A, and the drop is independent of the phase of
the observations.  This drop coincides with a drop in the X-ray flux
observed by \xte, where the baseline flux between flares ($\sim
100\;\mathrm{ct}\,{\mathrm s}^{-1}$\ for the All-Sky Monitor (ASM) on
\xte) began a slow decline around 2000 March (MJD 51600).  This
downward trend of the H$\alpha$ equivalent width is consistent with
the trend seen over the last 25 years (Paper I, Table~2), where the
observations in 1976 May found $\ew=580\A$, declining to
$120\A$\ in 1997.  Rough flux-calibration of this data and the
archival data shows that this change in the equivalent width is not
due to an increase in the brightness of the continuum, and therefore
must reflect a real and dramatic decrease in the brightness of the
emission line.

\subsection{He {\sc i} lines}
\label{sec:He-I-lines}

The spectra all show emission lines of \HeI\ $\lambda 6678$\ and
$\lambda 7065$.  In most cases, these lines are quite faint; the
equivalent width of the \HeI\ lines also dropped dramatically between
1999 and 2000.  We measured the equivalent width of the lines by
fitting Gaussians in the same manner as for the H$\alpha$\ lines.  For
most of the spectra, a single gaussian for each line was sufficient.
For the 1999 July 11 lines, the \HeI\ lines were double peaked just
like the H$\alpha$ line. The 1999 August 20 line required two
components for a satisfactory fit.

\begin{figure}
     \centerline{\psfig{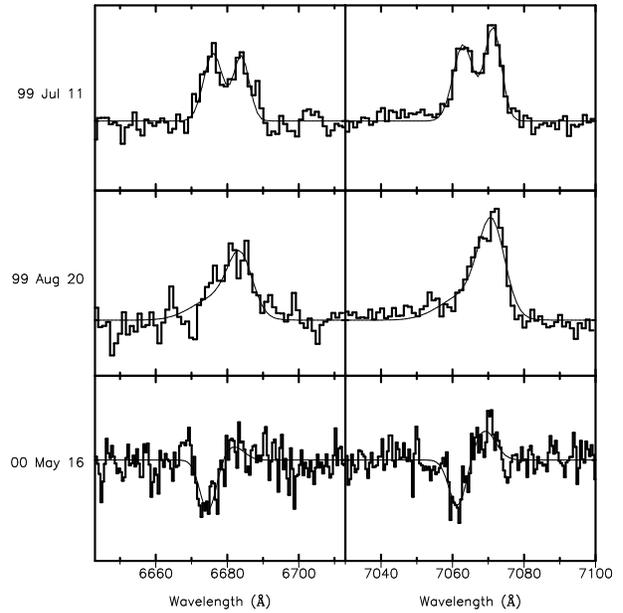}}
\caption{Selected line profiles of \HeI, showing the fits to the
  lines.  The spectra have been normalised by a polynomial fit to the
  continuum; the wavelengths of the fits were constrained to have the
  same velocity and width. The left side of the figure shows the
  $\lambda6678$\ line, the right side the $\lambda7065$\ line. The
  1999 July 11 spectrum shows a double-peaked profile, the 1999 August
  20 spectrum shows an asymmetric line, and the 2000 May 16 spectrum
  shows a P~Cygni profile.}\label{fig:HeIlines}
\end{figure}

The spectrum from 2000 May 16 clearly showed P Cygni profiles in both
the \HeI\ lines (Figure~\ref{fig:HeIlines}).  P~Cygni profiles were
recently discovered in X-ray spectra of \cir\ taken by {\em Chandra}
\cite{bs00}. The peak-to-valley distance of the profile in the optical
spectra is 7.62\A, or 340\kms.  This is a much smaller velocity
than that seen in the X-ray spectra (2000--3000\kms).

\begin{figure}
     \centerline{\psfig{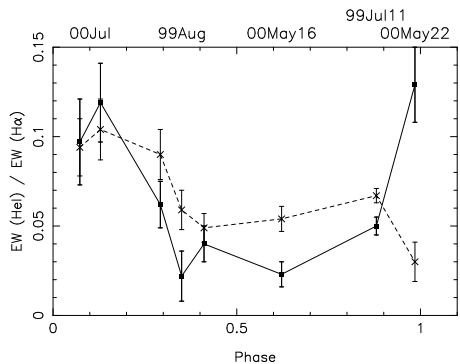}}
\caption{Equivalent width of the \HeI\ lines, normalised by the
  H$\alpha$\ equivalent width, as a function of phase. The filled
  squares show \HeI\ $\lambda$6678, the crosses \HeI\ $\lambda$7065.
  The dates of the observations are indicated above the plot.  The
  spectrum at phase 0.62 (2000 May 16) showed P~Cygni profiles in the
  \HeI\ lines: only the equivalent width of the emission component has
  been plotted here, thus underestimating the intrinsic line
  strength.}\label{fig:HeIew}
\end{figure}

Figure~\ref{fig:HeIew} shows the equivalent width of the \HeI\ lines
as a function of phase. In order to remove the dramatic secular
variation which occurred between 1999 and 2000, we normalised the
\HeI\ equivalent widths by the equivalent width of the H$\alpha$\ line
(Table~\ref{tab:fits}, column 3). The results show a smooth variation
with phase; both transitions are of similar strength, except for the
abrupt reversal in the phase 0 spectrum, when the $\lambda$7065
line becomes significantly weaker than the $\lambda$6678 line.

\subsection{Velocities of the emission lines}
\label{sec:Velocities}

The measured velocities of the emission line do not seem to conform to
any sensible pattern. In particular, there is no discernible signature
of orbital motion over most of the orbit. This implies that the site
of the emission is changing throughout the orbit, and possibly over
the one year spanning our observations -- see
Section~\ref{sec:discussion} for a discussion of the implications for
a model for the system.

\begin{figure}
     \centerline{\psfig{figure=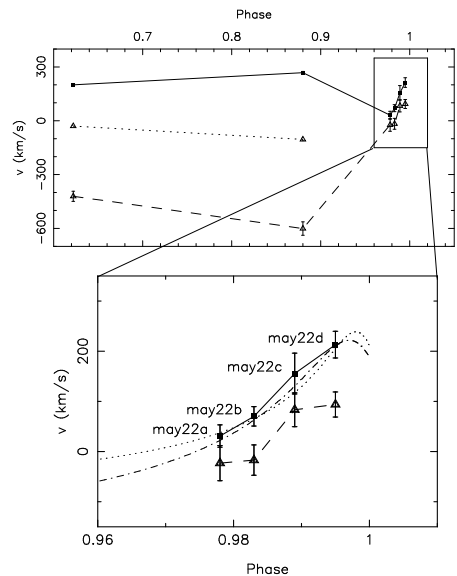,width=6cm,clip=t}}
     \caption{Velocities of the fitted components near phase 0.  The
       data from 2000 May 22 has been divided into four spectra of
       duration 7200~s; the rapid swing of velocity in both the narrow
       (solid line) and broad (dashed line) components is evident in
       the expanded section to the lower right.  The dotted line shows
       the second narrow (blueward) peak in the double-peaked spectra
       on 99 July 11 and 2000 May 16. The dotted and dot-dashed lines
       show two representative orbital curves, assuming $M_1 =
       1.5\Msolar$, $M_2 = 3.0\Msolar$; the dotted line has
       $i=45\deg$, $e=0.9$, while the dot-dashed line has $i=85\deg$,
       $e=0.87$; both curves have the longitude of periastron set to
       $210\deg$\ and a systemic velocity of $-120\kms$\ and $-220\kms$\ 
       respectively. }\label{fig:peri-v}
\end{figure}

However, we do see significant orbital motion in the data taken near
periastron. We divided the data taken on 2000 May 22 into four equal
segments of 7200~s duration and performed the same analysis on each
segment. The results are shown in Fig.~\ref{fig:peri-v}. Both
components of the H$\alpha$\ line, as well as the \HeI\ $\lambda$6678
line (not plotted), show an increase in velocity of 120\kms\ over six
hours.  This certainly implies that at this point in the orbit the
emission line is tracking the orbital motion; in a highly elliptical
orbit, nearly all the velocity swing happens right near periastron
\cite{tfh+99}. 

With only four velocity points, we cannot place any interesting
constraints on the orbital parameters. Two representative curves are
plotted with the observed points; it can be seen that there is
essentially no constraint on the orbital inclination, though the
eccentricity must be high to produce the observed sharp swing in
velocity.

\subsection{Timing of periastron}
\label{sec:Timing-periastron}

It is important to note that the phases have been calculated using the
ephemeris of Stewart et al. \shortcite{snp+91}. Phase 0 refers to the
onset of radio flares, and how this relates to the X-ray dips observed
by \xte\ is not clear.  We have examined the quick-look results
provided by the RXTE/ASM team to see if we can compare the observed
times of the X-ray dips with the predicted times from the Stewart et
al.  ephemeris.  The periastron passage on 2000 May 22 did not have
good ASM coverage; both the preceding and the following X-ray dips
appear to have occurred 0.4--0.5~d later than the phase 0 predicted by
the Stewart et al. ephemeris, corresponding to a shift of 0.024--0.03
in phase.  Given that the ephemeris was based on radio flares, and
that the X-ray light curve changes so dramatically, this agreement is
reasonably good.  In future work we hope to further investigate the
exact relationship between the X-ray, radio and optical behaviour.

\section{Discussion}
\label{sec:discussion}

\subsection{A model for the system}
\label{sec:model-system}

In Paper~I, we proposed a theoretical model for Cir~X-1. The binary is
a low-mass neutron-star binary with an ultra-eccentric binary orbit
($e \ga 0.7$, possibly as high as 0.9; Tauris et al.
1999\nocite{tfh+99}).  During periastron passage, the companion star
(which is a subgiant of about 3--5\Msolar) overfills its Roche-lobe,
causing a transfer of mass at a super-Eddington rate, which in turn
drives a strong matter outflow.  After periastron, mass transfer from
the companion ceases, but accretion continues at a near-Eddington rate
as the neutron star captures the residual matter in its Roche-lobe. An
accretion disk gradually forms, which is initially non-Keplerian and
geometrically thick. This change between quasi-spherical and disk
accretion causes the change between strong X-ray variability after
phase 0.  Between phases 0.5--0.9, the accretion disk becomes
Keplerian and there is steady accretion. The disk is then disrupted by
tidal forces during the following periastron passage.

Based on the consistently high velocities seen in all the data up to
that time, we suggested in Paper~I that \cir\ might have a very high
radial velocity.  Our new spectroscopic data do not show evidence of
high line-of-sight velocities, though the argument about the velocity
from the association with G321.9$-$0.3 is still valid.

\subsection{Sites of emission}
\label{sec:Sites-emission}

Our data and the archival data show a persistent broad component in
the H$\alpha$ line, indicating the presence of a high-velocity flow.
One possibility for the line-formation region is the accretion flow
close to the neutron star.  The X-ray luminosity of Cir X-1 reaches
the Eddington limit at the phases shortly after periastron. This
implies a mass accretion rate $\dot M \ga 2 \times 10^{18}\,{\mathrm
  g}\,{\mathrm s}^{-1}$.  The quantity $n_e R^2$ is given by
\begin{eqnarray}
  n_{\rm e} R^2 &\approx& 1.0 \times 10^{33} 
  \left( \dot M \over 2\ee{18}\gs \right) \nonumber\\
  & & \times \left( 1000\kms \over v \right) \chi^{-1} \;{\rm cm}^{-1} \ , 
\end{eqnarray}
where $n_{\rm e}$ is the electron number density. $\chi$ parametrises
the degree of spherical accretion, and its value is unity for
spherical accretion.  Hence the ionisation parameter
\begin{equation}  
  \xi \equiv  {L_{\rm x} \over {n_e R^2} } \sim 10^5 \chi^{-1}
  \qquad(\mbox{in cgs units})  \ ,
\end{equation}
implying a temperature $T \sim 10^7\;$K for the accreting matter (see
Hatchett et al. 1976\nocite{hbm76}).  The strong X-ray emission from
the source would therefore completely ionise the accretion gas with
inflowing velocity with $v \sim 1000\kms$, making it unlikely to emit
H Balmer lines.
 
Alternatively, the broad component of the H$\alpha$\ line could be
emitted from the outflow matter. A characteristic of a radiatively
driven outflow is that the outflowing matter is accelerated from a
slow initial velocity and eventually reaches a terminal velocity. As
there is a negative gradient from the observer to the outflow matter,
the blue wing of the line suffers very strong self-absorption. The
line would therefore show a P~Cygni profile. Since our optical data do
not show the P~Cygni profile typical of such a radiatively driven
outflow (with the exception of one set of spectra: see below), the
gradient of the line-of-sight velocity in the line formation region
must be positive, i.e. the outflow is decelerating. If the broad
component originates from a high-velocity outflow, the line formation
region is in the decelerating zone, and the accelerating zone is
behind the photosphere. Such outflow might be caused by an ejection of
high-velocity material during the violent mass transfer epoch at the
periastron.

On one occasion, we do see an absorption component in the blue wing of
the \HeI\ lines, though not in the H$\alpha$\ line. We suggest that,
instead of arising in a radiatively driven outflow, we are seeing
absorption from the outflowing material far from the system,
superimposed on emission lines from the accretion disk.  The absorbing
matter consists of material ejected from the system in previous
episodes of violent mass transfer, and the absorption is occurring at
some distance from the system, where the outflow has decelerated
significantly from the 2000--3000\kms\ seen in the X-ray spectra.

The absorption is seen only in spectra from 2000 May 16, which is when
the \HeI\ emission from the disk was weakest (Table~\ref{tab:fits}).
We suggest that the absorption in the outflow is masked in the other
spectra by the strength of the emission line, which is also why we do
not see similar absorption in the H$\alpha$\ line.


\subsection{First detection of an accretion disk?}
\label{sec:Detection-accretion-disk}

Our detection of the double-peaked lines formed in an accretion disk
is strong evidence for the eccentric binary model for \cir. The
double-peaked lines were seen twice, in spectra taken almost a year
apart, at phases 0.88 and 0.62. Spectra taken in the first half of the
orbit all show an asymmetric H$\alpha$\ emission line, while the
spectrum taken at periastron (phase 0.98) shows a completely symmetric
line.

Double-peaked lines are often found in cataclysmic variables (e.g.\ IP
Peg, see Marsh 1988\nocite{mar88}), which are binaries with a white
dwarf accreting material from a low-mass companion star, and in
black-hole binaries during the high-soft state (e.g.\ GX339$-$4, see
Soria, Wu \& Johnston 1999\nocite{swj99}) or quiescence (e.g.\
A0620$-$00, Johnston, Kulkarni \& Oke 1989\nocite{jko89}).  They are
also found in the accretion-disk corona sources (e.g.\ 2A~1822$-$371,
Mason et al. 1982\nocite{mmt+82}).  They are, however, rare amongst
other X-ray binaries, and X-ray pulsars do not typically show
double-peaked optical lines\footnote{Though note that 2A~1822$-$371
  was recently shown \cite{jk01} to be an X-ray pulsar!}.

It was shown \cite{wshj01} that double-peaked lines can be formed in a
temperature-inversion layer on an opaque accretion disk when the disk
is irradiated by soft X-rays. The X-ray spectrum of Cir X-1 is soft,
dominated by a black-body component with an effective temperature $T
\sim 2\,$keV \cite{slb99}. Thus the double-peaked lines observed in
Cir X-1 probably indicate the presence of an accretion disk.
Moreover, a significant part of the disk surface is not obscured from
the central X-ray source so that the irradiatively induced temperature
inversion can occur.

It is interesting to note that the 1999 July 
(Fig.~\ref{fig:Halpha-fits}, phase 0.88) observation, which was the
first spectrum to show double-peaked lines, was taken in the cycle
after an orbit during which \cir\ did not have a post-periastron X-ray
flare; only a handful of orbits observed by XTE had not shown such a
flare before then.  It is possible that this has some bearing on the
visibility of the optical lines from the accretion disk, for instance
if the absence of an X-ray outburst means that matter which usually
obscures the disk was more transparent than usual. Alternatively, the
accretion rates during this observational epoch were low enough to
allow a mature Keplerian disk to develop within an orbital cycle.

The double-peaked lines seen in 2000 May (Fig.~\ref{fig:Halpha-fits},
phase 0.62) did not occur after such an anomalous orbit.  However, by
that stage the X-ray flux observed by \xte\ had begun to drop (see
Sec~\ref{sec:Secular-variation}, Fig.~\ref{fig:eqw}), which might also
imply that veiling material was thinner.

It is possible that, instead of being an intrinsically double-peaked
line, the profile observed on 2000 May 16 is actually flat-topped.
Such lines have seen in several black hole candidates, such as
GX~339$-$4 \cite{sfl99,swj99} and GRO~J1655$-$40 \cite{swh00}, where
they are interpreted as arising in a horizontal wind launched from a
disk \cite{mc97}. Possibly the line in \cir\ arises in a similar
fashion; however, this requires the emission site to change from a
spherical outflow (phases $< 0.5$; Fig.~\ref{fig:Halpha-fits}) to a
horizontal wind (phase 0.6) to a disk (phase 0.88). The interpretation
of the line as a barely-resolved double peak has at least the virtue
of being a simpler explanation.

\begin{figure}
     \centerline{\psfig{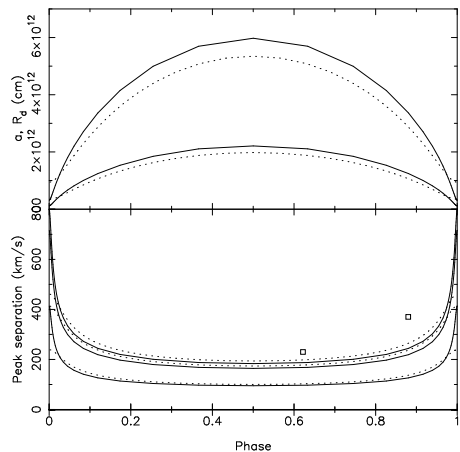}}
     \caption{Sketch of the separation of the double-peaks as a
       function of orbital phase, for some representative orbital
       parameters, compared with our observed values.  The top panel
       shows the separation between the two stars (top line) and the
       outer radius of the disk, where we assume the disk is limited
       by the size of the Roche lobe of the compact star and its outer
       radius is truncated at $0.7\;R_L$.  The masses of the two stars
       are $M_1 = 1.5\Msolar$, $M_2 = 3.0\Msolar$, and the orbital
       eccentricity and inclination are $e = 0.9$\ (solid lines) and
       $e=0.7$\ (dotted lines). The bottom panel shows the predicted
       peak separation for such a disk, for a range of orbital
       inclinations: (from the bottom) $i=30\deg$, $45\deg$\ and
       $75\deg$.  The open squares show the measured peak separation
       for the two dates on which they were observed.
       }\label{fig:disksize}
\end{figure}

Assuming the double-peaked lines arise in an accretion disk, then the
separation reflects the velocity at the outer edge of the emission
region of the accretion disk \cite{sma81}. The separation of the peaks
was very different on the two occasions the double-peaked profile was
observed (370\kms\ in 1999 July, 230\kms\ in 2000 May), which suggests
that the accretion disk had a very different size.  This could be due
to the different phase of the observation, implying the disk is
shrinking as periastron approaches. Figure~\ref{fig:disksize} shows a
comparison between the observed peak separations and a model where the
size of the disk is truncated by the size of the Roche lobe in an
eccentric orbit. This represents the largest possible size for the
accretion disk, and hence a {\em lower} limit to the separation of the
two peaks. The fact that the observed velocities are above this limit
implies that either the disk has not reached the truncation radius, or
that the emission region for H$\alpha$\ is not at the outer edge of
the disk. The former would imply that, after the disk is disrupted at
periastron, it has not had time to grow to the truncation radius
before the Roche lobe begins to contract again
(Fig.~\ref{fig:disksize}), which would imply that the viscous
timescale for the growth of the disk is longer than the dynamical
timescale, which is 8~d for half an orbit. 

Alternatively, the change in peak separation could be the result of a
secular variation in the size of the emitting region; or the
temperature of the disk could be varying, so that the location of the
H$\alpha$\ emission site is changing. In this case, the smaller peak
separation implies a hotter disk, with the emission site having moved
towards the outside of the disk.

Other explanations for the double-peaked profiles are possible which
do not require the presence of an accretion disk. They could, for
instance, arise in a bipolar jet, since a relativistic jet has been
inferred from high-resolution radio maps of the source \cite{fst+98}.
However, in SS~433, which has a jet with velocity $\sim 0.3c$, the
separation of the components is hundreds of {\aa}ngstroms
\cite{mar84}, compared to the 5--8\A\ seen in our spectra. Such a
small separation could only be produced by a relativistic jet if the
jet was almost in the plane of the sky (in which case the transverse
Doppler shift would cause both components to be redshifted). Thus any
jet would need to be non-relativistic to produce the observed lines.

Variable double-peaked lines have also been seen in infrared spectra
of Cyg~X-3 \cite{fhp99}, where they were interpreted as arising in a
disk-like wind outside the binary orbit. Such a disk-wind requires a
large angular momentum, which would be unlikely in a system like \cir\ 
with such a large orbital period.  The double-peaks could arise from
emission in different parts of the system; however, we note that the
standard sites for such localised emission, such as the hot spot where
the accretion stream impacts the disk, still require the presence of a
disk in the system.

Finally, it is possible that the central dip in the double-peaked
profile is formed by absorption in front between us and the emitter,
especially given the fact that we see absorption in the \HeI\ lines on
a different occasion (see \S~\ref{sec:He-I-lines}).  Such lines are
seen in symbiotic stars, such as CH~Cygni, which shows two symmetric
peaks separated by a deep central reversal \cite{aon80}; in this
object, the profile is assumed to arise in a self-absorbed wind.
However, in the case of the profiles seen in \cir, the conditions for
such absorption would be hard to produce. Since the profile is
symmetric, the absorbers would need to be at a very similar velocity
to the emitters; and since the central dip is broad and does not reach
the continuum, the absorbers would need a wide velocity dispersion and
only a small optical depth. The most probable interpretation of the
profile is that it arises from the Doppler motions in an accretion
disk.

\subsection{Long-term changes in the emission lines}
\label{sec:Changes-emission}

The brightness of the emission lines has decreased dramatically over
the past 25 years, from $\ew=580\A$\ in 1976 to $25\A$\ in
2000.  Our interpretation is that the contribution of different
emission sites to the emission lines has been shifting over the years.
The contribution from the accretion disk is likely to have been
essentially constant, since the temperature of the disk is limited by
the accretion rate (assumed to be at Eddington) and the size of the
disk is limited by the size of the orbit. The contribution to the
emission line from other regions -- the outflow, the companion star --
can change dramatically depending on the state of the system. It is
possible that we have seen the disk for the first time precisely
because the contributions from other sites in the orbit have lessened,
so the H$\alpha$\ line is now dominated by emission from the accretion
disk.

\section*{Acknowledgments}
\label{sec:Acknowledgments}

We thank the ANU RSAA Time Assignment Committee for their generous
allocation of time to this project. We thank Roberto Soria for
assistance with the observations.  We also thank the referee for
providing useful suggestions and clarifications, particularly for
suggestions about the flat-topped profile.  KW acknowledges the
support from the ARC through an ARC Australian Research Fellowship.
This work is partially supported by an ARC/USyd Sesqui R\&D Grant.


\end{document}